\documentclass{elsarticle}
\usepackage{a4,hyperref}

\begin{document}

\begin{frontmatter}

\title{Interface shapes in microfluidic porous media: conditions allowing steady, simultaneous two-phase flow}

\author{S.J. Cox}
\address{Department of Mathematics, Aberystwyth University, Ceredigion SY23 1HH, UK}

\author{A. Davarpanah}
\address{Department of Mathematics, Aberystwyth University, Ceredigion SY23 1HH, UK}

\author{W.R. Rossen}
\address{Delft University of Technology, Delft, 2628CN / 2600GA, Netherlands}

\begin{abstract}

Microfluidic devices offer unique opportunities to directly observe multiphase flow in porous media. However, as a direct representation of flow in geological pore networks, conventional microfluidics face several challenges. One is that simultaneous two-phase flow is not possible in a two-dimensional network without fluctuation occupancy of pores. Nonetheless, such flow is possible in a microfluidic network if wetting phase can form a bridge across the gap between solid surfaces at a pore constriction while non-wetting phase flows through the constriction. We call this phenomenon ``bridging".

Here we consider the conditions under which this is possible as a function of capillary pressure and geometry of the constriction. Using the Surface Evolver program, we determine conditions for stable interfaces in a constriction, the range of capillary pressures at which bridging can occur, and those where the wetting phase would invade and block the constriction to the flow of the non-wetting phase (``snap-off").

We assume that the channels have uniform depth, vertical walls, and flat bottom and top surfaces, and that one phase perfectly wets the solid surfaces. If the constriction is long and straight, snap-off occurs at the same capillary pressure as bridging. For long, curved channels, snap-off happens as liquid imbibes before bridging can occur. For constrictions between cylindrical pillars, however, there is a range of capillary pressures at which bridging is stable; the range is greater the narrower the diameter of the cylinders relative to the width of the constriction. For smaller-diameter pillars, the phenomenon of ``Roof" snap-off, that is, snap-off as non-wetting phase invades a downstream pore body, is not possible.

We relate these results to the shape of pore networks commonly used in microfluidic studies of two-phase flow to consider whether two-phase flow is possible in these networks without fluctuating pore occupancy.
\end{abstract}


\end{frontmatter}

\clearpage

\section*{Graphical Abstract}

\begin{figure}[h]
\centerline{
\includegraphics[angle=0,width=\textwidth]{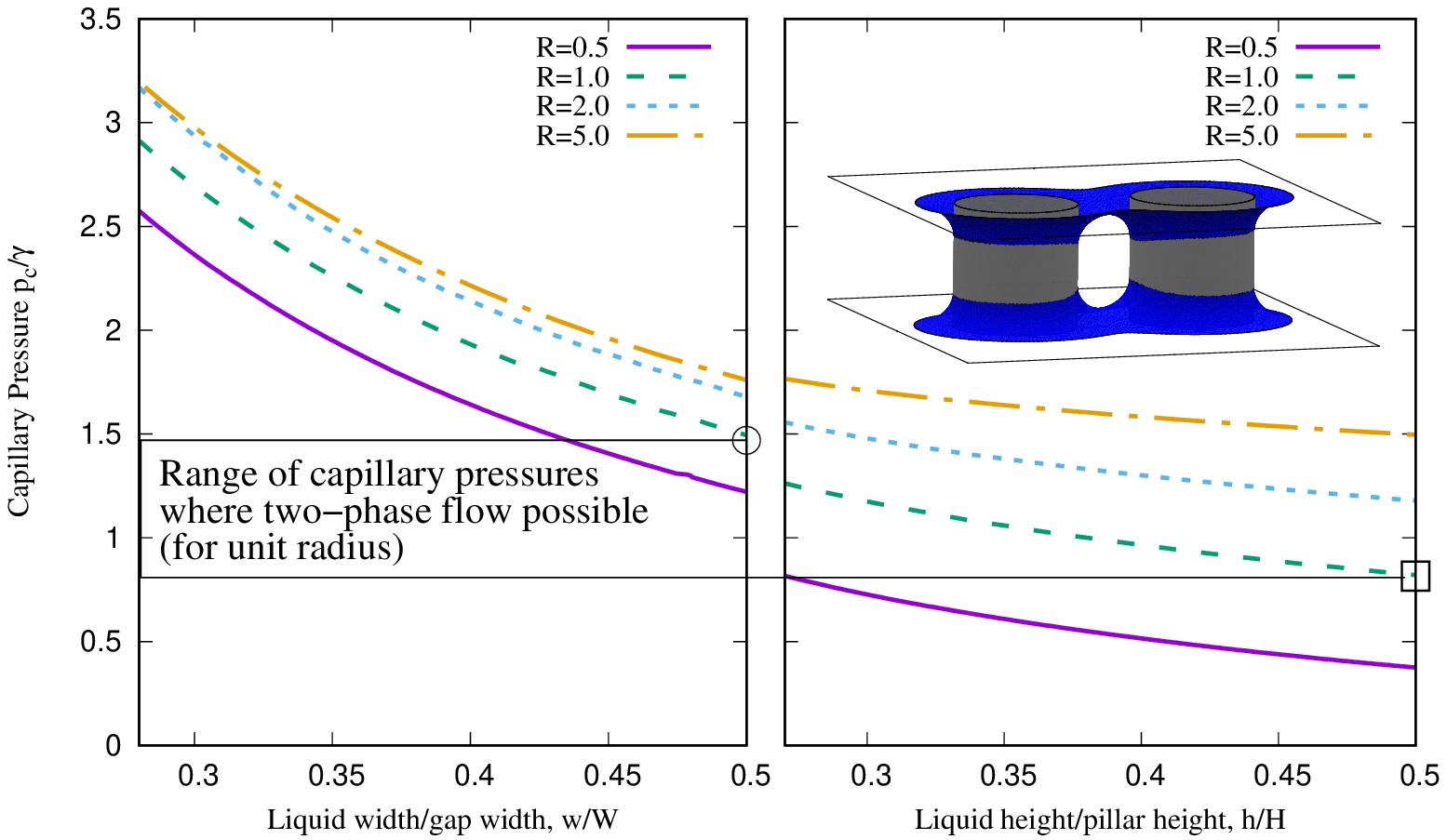}
}
\end{figure}

\clearpage

\section{Introduction}

Microfluidic systems, or micromodels~\cite{haghighi94}, are useful for the study of flow in subsurface (geological) porous media~\cite{cuib17}. They have the distinct advantage that one can see the flow dynamics directly, allowing observation of interfacial interactions, mass-transfer processes, phase behaviour and wetting transitions, for example. Such insight is likely to be of use in determining the dynamics of fluids in  a range of applications such as oil recovery, aquifer remediation, and carbon capture and storage.

\begin{figure}
\centerline{
\includegraphics[angle=0,width=0.7\textwidth]{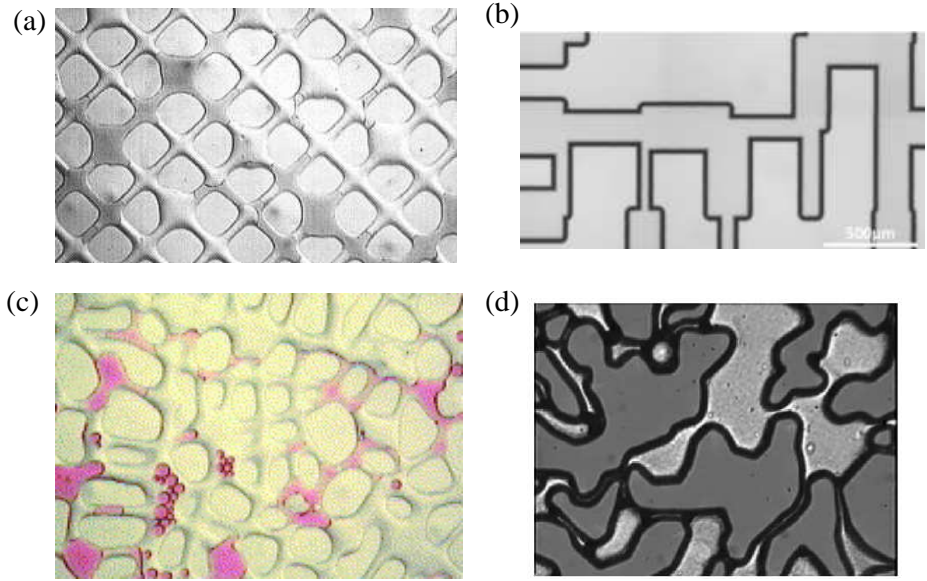}
}
\caption{Examples of microfluidic network representing an idealized porous medium. (a) Highly ordered (from ~\cite{jeong03}). (b) Disordered artificial network (from~\cite{jeong03}). (c) Disordered with rectangular pore network (from~\cite{marchandbks17}.  (d) Network based on a cross-section of rock (from~\cite{kovscektr7}).}
\label{fig:network}
\end{figure}

\begin{figure}
\centerline{
(a)
\includegraphics[angle=0,width=0.6\textwidth]{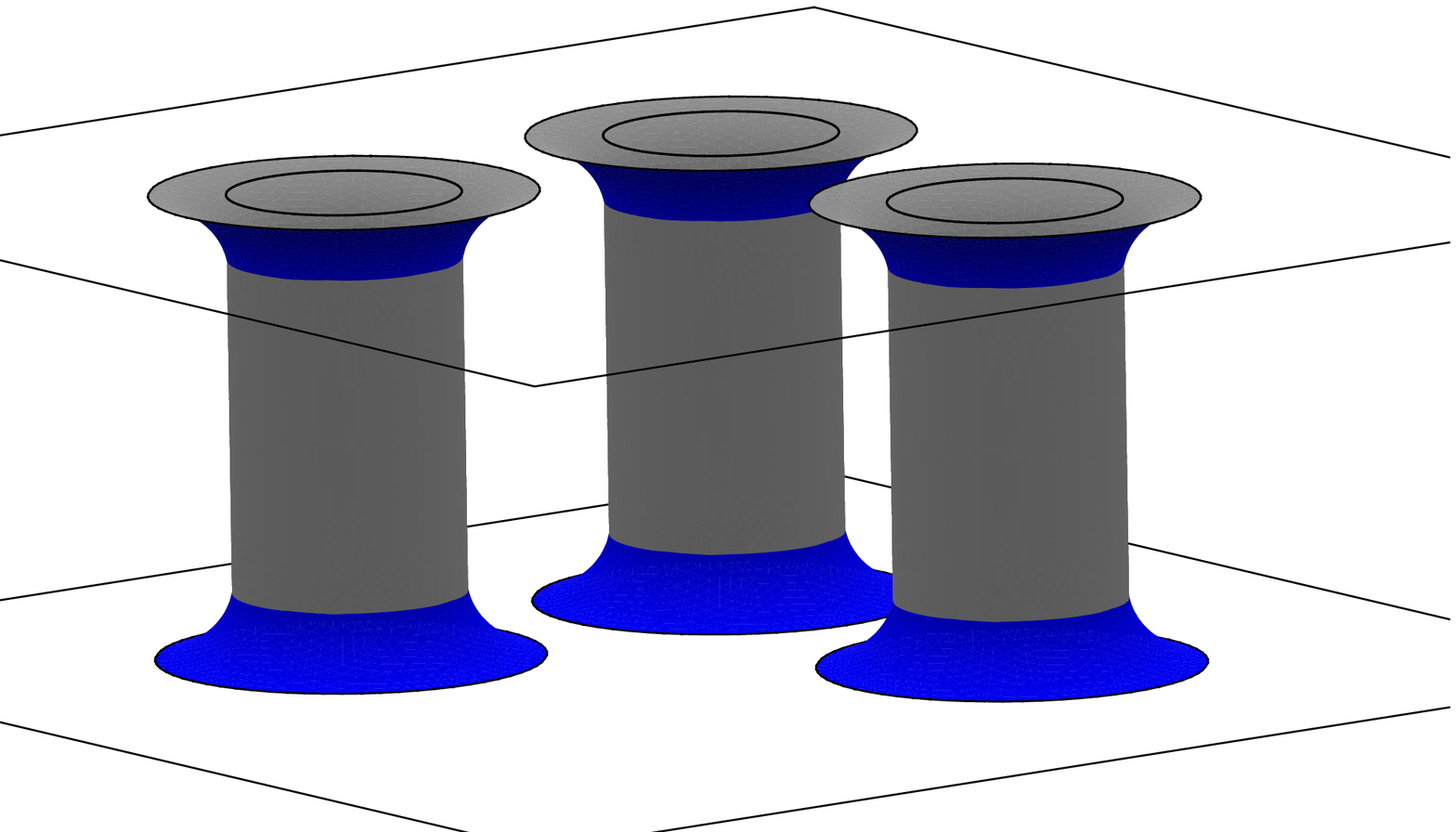}
(b)
\includegraphics[angle=0,width=0.4\textwidth]{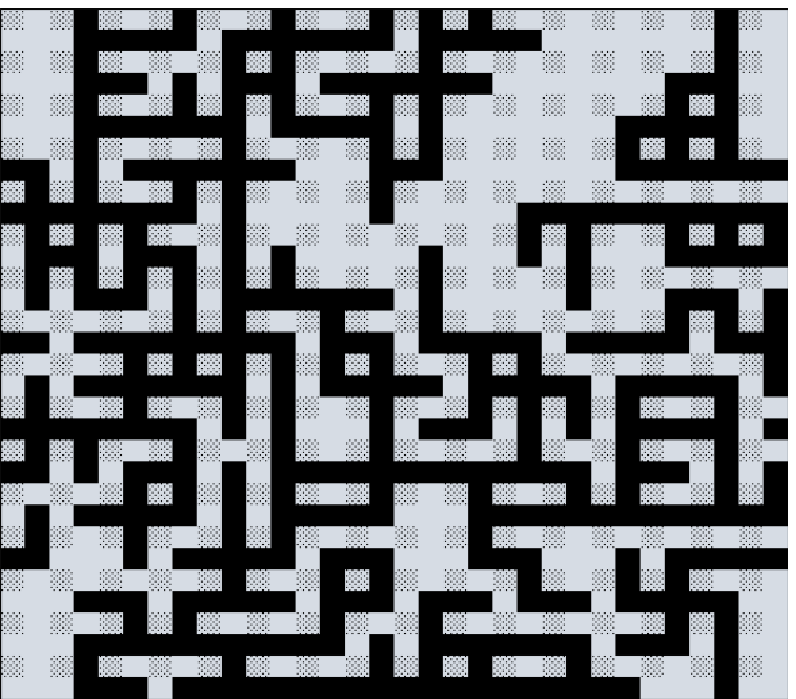}
}
\caption{(a) Wetting phase surrounds the bases and tops of pillars between parallel plates, while non-wetting phase occupies the interior. (b) Example of invasion percolation on a network of coordination number four, from~\cite{holstvoogd20}. Black is NWP, gray squares represent pillars (which could have any shape), pale gray is WP. Periodic boundary conditions apply to top and bottom, and left and right edges. Capillary pressure is sufficient for NWP to enter 50\% of pore throats. This is just past the point where NWP can flow from both top to bottom and left to right. Note there is no path for WP flow from either top to bottom or left to right without bridging.
}
\label{fig:simulated_network}
\end{figure}

A microfluidic porous medium usually has a geometry consisting of channels of uniform depth bounded by vertical walls. Viewed from above, as in the examples in figure~\ref{fig:network}, the network can have nearly any (2D) shape desired. We call solid barriers to flow (defined by vertical walls) ``pillars", and consider two-phase flow consisting of a non-wetting phase (NWP) which, in those pores that it occupies, occupies the interior of the channels, and a wetting phase (WP) that occupies the corners where pillars meet near the top and bottom surfaces (figure~\ref{fig:simulated_network}(a)). In a complex network of channels there will be narrower and wider regions: we call the narrow locations ``pore throats'' and the wider locations ``pore bodies". Multiple pore throats may connect to one pore body in a network.

Despite the relative simplicity of a 2D microfluidic network, there are many similarities to 3D pore networks in geological formations, making them a useful proxy for predicting flow characteristics there. These include non-uniform channels/pores, and in particular the presence of pore throats, where capillary forces are significant. 
Moreover, it is straightforward to control and adjust the wettability, e.g. by treating surfaces/coating walls with specific minerals, and to capture a wide range of possible pore geometries. It is also possible to get multiple 3D layers of pores using 3D printing; we do not discuss such networks further here.

Nonetheless, care is needed in extrapolating directly from flow in microfluidics to flow in other porous media. In particular, the channel depth is uniform in most current designs of microfluidic channels (figure~\ref{fig:network}), and so the expansion in the channel from a pore throat into a pore body is less extreme  than in geological porous media. In rheological terms this tends to emphasize shearing over elastic contributions to the viscosity. This also often means that the widths observed from above are not the most important widths for estimating capillary forces.

Moreover, the network itself is two-dimensional (2D), although the channels are 3D of course. In a 2D network, simultaneous, steady two-phase flow is not possible without fluctuating occupancy of pores~\cite{fisher61}. Simultaneous two-phase flow {\em is} possible in 3D geological networks, without fluctuations, which is an essential aspect of flow in those networks~\cite{kingm19}. In a microfluidic network, however, it is possible for the wetting phase to cross a throat at top and bottom and connect wetting phase residing in corners around two pillars. We call this a ``bridge": it means that the flow paths of WP and NWP can cross in a microfluidic device, unlike in a 2D network.

Suppose that a microfluidic network is initially full of the wetting phase, and then we co-inject WP and NWP at fixed fraction. Initially the NWP cannot flow across the network and its saturation (volume fraction in the pore space) rises until it can form a continuous path through the network, as illustrated in figure~\ref{fig:simulated_network}(b). At this point the capillary pressure $p_c$ is too high for bridging, and the WP has no path for flow through the network. The WP therefore accumulates, and the capillary pressure falls, until either bridging occurs at enough throats that WP can flow at the injected fractional flow, or pore occupancy starts fluctuating between WP and NWP. In this article we determine the conditions under which bridging of the wetting phase is possible.

\begin{figure}
\centerline{
\includegraphics[angle=0,width=0.8\textwidth]{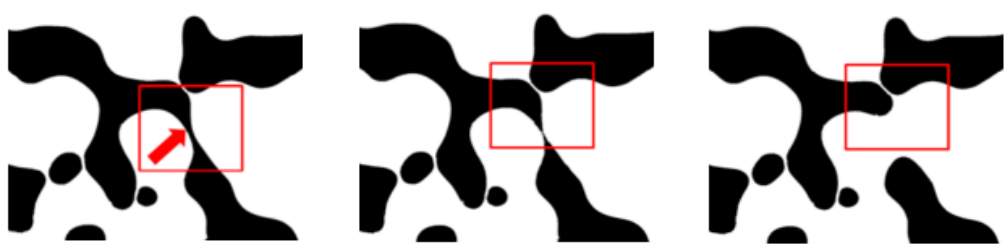}
}
\caption{A pore structure in which gas (black) and liquid (white) are co-injected, leading to snap-off and the formation of a bubble. The images show three time-steps from an experiment, at 0, 1.23 and 1.63 seconds. Adapted from \cite{alquaimir19}.}
\label{fig:snapoff}
\end{figure}

\begin{figure}

\centerline{
(a)
\includegraphics[angle=0,width=0.3\textwidth]{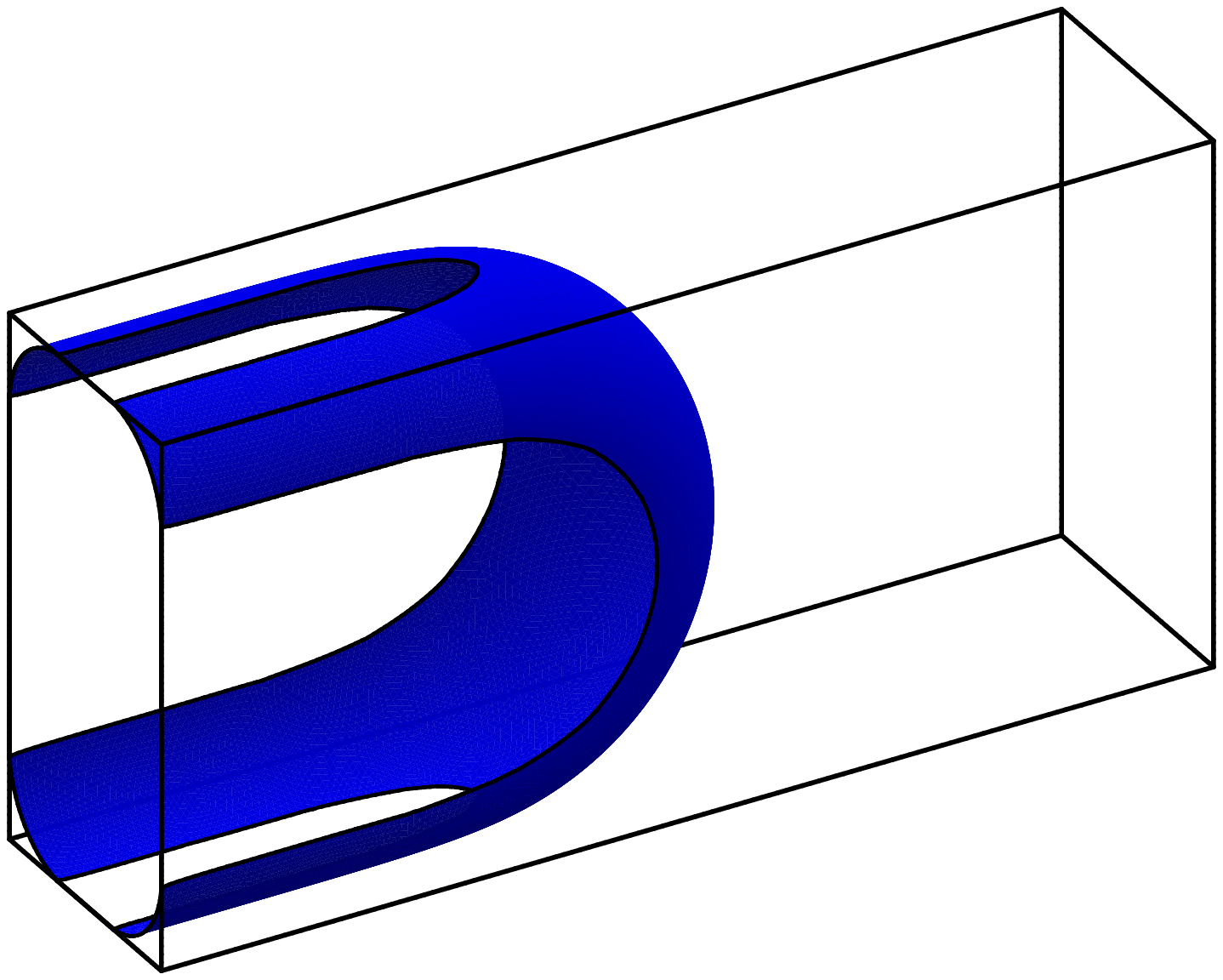}
(b)
\includegraphics[angle=0,width=0.3\textwidth]{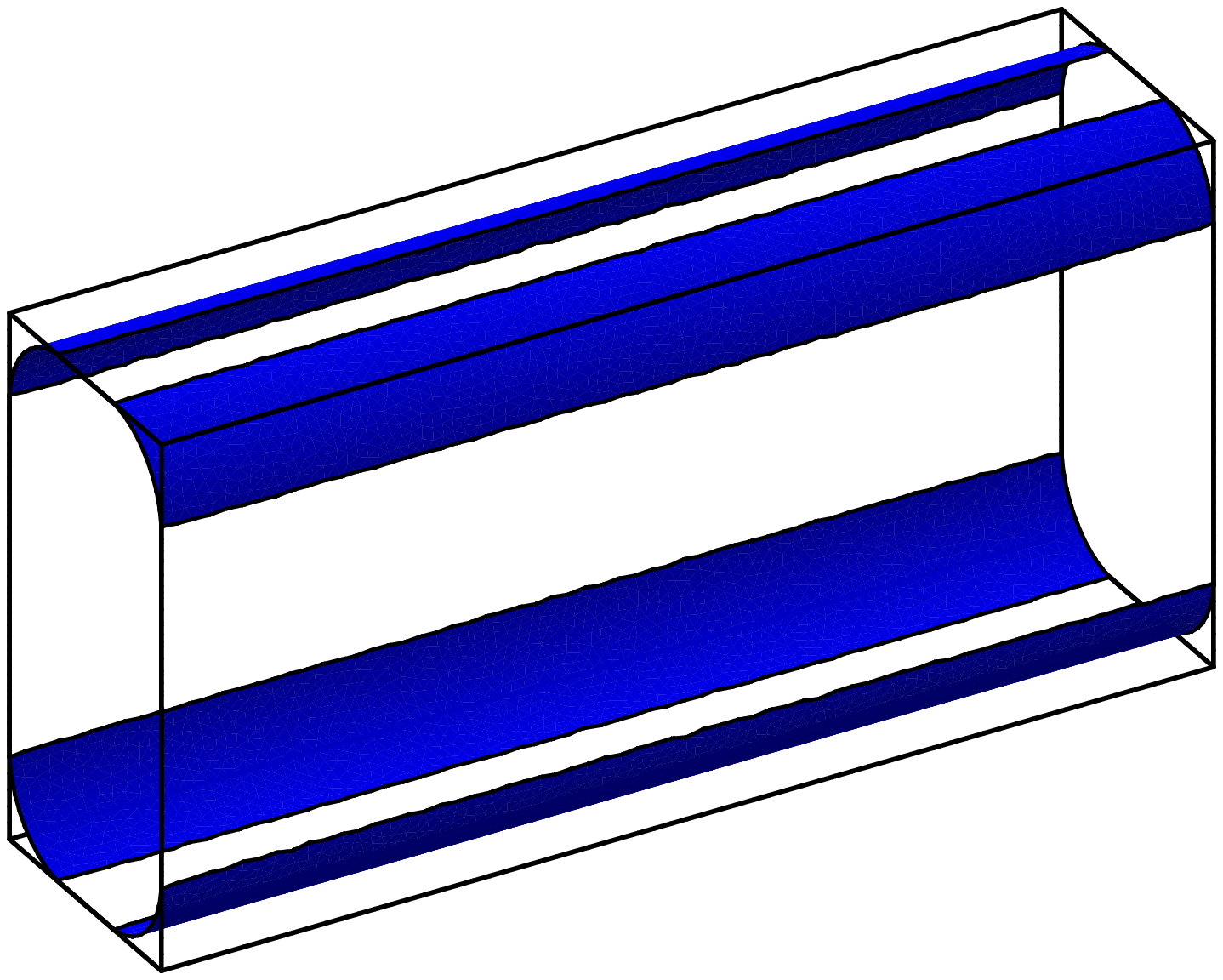}
}
\caption{(a) The interface separating NWP from WP moves from left to right along a straight rectangular channel after overcoming the capillary entry pressure.  (b)  After the front has left the channel, it leaves behind four narrow regions of the WP.  The wetting films on the walls are not shown. The channel has height $H=2$ and width $W=1$ and the contact angle is $2^\circ$. }
\label{fig:straight}
\end{figure}

Consider a long, straight pore throat with a uniform rectangular cross-section, in which WP lines the corners of the throat (as in figure~\ref{fig:straight}(b) below). The interfaces between the phases  are cylindrical and make contact when the diameter is equal to the width of the channel. That is, bridging can occur if the throat is at least as tall as it is wide. As the two interfaces then approach each other in the centre of the throat there is no change in interface curvature. The condition on the capillary pressure $p_c$ for bridging to occur is, therefore, the same as the condition on $p_c$ for WP to swell and block the whole throat. 
This is what we call ``snap-off", as illustrated in Figure~\ref{fig:snapoff}.
In other words, once bridging occurs, any slight reduction in the capillary pressure in the surrounding medium would cause snap-off in the throat; any slight increase in capillary pressure would cause the bridge to disconnect.

If, instead, the pore throat is not straight, but passes between curved walls (either with the same or opposite curvatures) then the curvature of the pillar walls affects the conditions for both bridging and snap-off. In this study we examine conditions (specifically, the range of capillary pressure) for which bridging is possible without snap-off, as a function of throat geometry. We then discuss the implications for the feasibility of two-phase flow without fluctuating pore occupancy in microfluidic devices: that is, the extent to which microfluidic devices are able to represent this aspect of multiphase flow in geological porous formations under capillary-dominated conditions.

Before it can flow, NWP must first overcome a capillary {\it entry} (or threshold) pressure $p_c^e$. LeNormand et al.~\cite{lenormand} predict that the capillary entry pressure in a rectangular channel of cross-sectional width $W$ and height $H$ is
\begin{equation}
 p_c^e = \gamma \left( \frac{2}{W} + \frac{2}{H} \right),
\label{eq:lenormand}
\end{equation}
where $\gamma$ is the interfacial tension. In the following we mostly consider the case $H=2$ and $W=1$, for which eq.~(\ref{eq:lenormand}) predicts $p_c^e/\gamma = 3$. 

The NWP then moves downstream, displacing WP (known as drainage~\cite{chen85,paterson84}), with the interface at the leading edge moving steadily along the channel; see figure~\ref{fig:straight}(a). In the corners behind this moving front, thin triangular regions of WP remain. The volume of these regions is determined by the curvature of the interface, which itself depends on the aspect ratio of the channel.

Once the front has fully-penetrated the channel, just the thin regions of WP in the corners remain; see figure~\ref{fig:straight}(b). Subsequent drainage or imbibition then causes these regions to shrink or enlarge, respectively. In a straight channel, all four WP regions in the corners have identical cross-section and are uniform along the channel. Ma et al.~\cite{mamorrow} relate the capillary pressure of these regions to the cross-sectional area $A_s$ of each region in terms of the contact angle $\theta$:
\begin{equation}
 p_c =  \frac{2 \gamma}{\sqrt{A_s}}  \sqrt { \sqrt{2} \cos(\theta)\cos\left(\theta+45^\circ\right)- \frac{\pi}{2}\left(1 - \frac{45^\circ+\theta}{90^\circ}\right)}.
\label{eq:mamorrow}
\end{equation}

If, upon subsequent imbibition, the corner channels swell sufficiently, they touch: if this occurs on the bottom or top of the channel, we refer to it as horizontal bridging. If it occurs on the sides of the channel we refer to it as vertical merging. Vertical merging does not enable WP flow across the throat.

We use the Surface Evolver software~\cite{brakke92}  (see \S~\ref{sec:method}) to predict the shape of the interfaces in various geometries by minimizing the surface energy of the interface between the two fluids to give accurate measurements of the capillary pressure $p_c$ (the pressure difference across the interface, calculated by the program). By making small changes in the NWP volume and repeating the minimization, we predict the quasistatic variation of pressure during imbibition or drainage. As well as neglecting viscous losses, we assume that the effects of gravity are negligible (small Bond number) on the scale of the channels.

We establish a benchmark for our simulations in a straight rectangular channel, determining the effect of the choice of channel aspect ratio and the contact angle at which the interfaces meet the walls of the channel (\S~\ref{sec:straight}). We then modify this channel to allow both side walls (pillars) to have curvature: either in the same direction, giving a curved rectangular ``duct'' (\S~\ref{sec:curved}), or in opposite directions, giving a flow between cylindrical pillars (\S~\ref{sec:pillars}) into a large pore body downstream.

The channel height $H$ is uniform in most microfluidic devices. We report capillary pressures in units of $\gamma/(H/2)$, the capillary entry pressure of a Hele-Shaw cell, i.e. a gap between two smooth, parallel plates. This is the capillary entry pressure of a pore body of extremely wide diameter in a microfluidic device. 
For a surface tension of $\gamma = 3\times 10^{-2} {\rm N/m}^2$ and a channel height $H = 25 \mu{\rm m}$, we have that a value of $p_c=3$ in our calculations described below corresponds to a capillary pressure of $7,200$ Pa.

The trends in capillary (entry) pressure with varying $H$ and $W$ are captured well by eqs.~(\ref{eq:lenormand}) and~(\ref{eq:mamorrow}), and so here we concentrate on the variation with changes in wall curvature.

\section{Simulation Method}
\label{sec:method}

The Surface Evolver program minimizes surface energy by moving WP and NWP, subject to a fixed solid geometry and fixed phase volumes. In this case the surface energy is the surface area of the interfaces multiplied by their surface tensions. Each interface is tessellated into many small triangles to allow it to curve; we generally use four levels of refinement (i.e. recursive splitting of one triangle into four smaller triangles) to provide an acceptable level of accuracy without simulations taking more than a few hours to explore all relevant parameter values.

The software explicitly accounts for the surface energies of each phase against solid and also against each other; by adjusting the solid/fluid phase surface energies, we can in effect fix the contact angle. We choose an interfacial tension $\gamma=1$ between WP and NWP and a tension $\gamma_w$ where the NWP touches the solid walls, which sets the contact angle from $\theta = \cos^{-1}(\gamma_w/\gamma)$. In the absence of gravity the value of the interfacial tension is not important, only the ratio $\gamma/\gamma_w$.

The case of perfect wetting - a contact angle of $0^\circ$ - is numerically slow to converge, so we prefer a small contact angle of a few degrees which ensures that simulations converge quickly and accurately. Figure~\ref{fig:data_straight}(a) shows that the prediction of capillary pressure with a contact angle of $2^\circ$ is almost indistinguishable from perfect wetting. 

At the entrance and exit of the channel, treated as vertical planes, we set the contact angle to $90^\circ$. This in effect gives a plane of symmetry to the fluid interface at the entrance and exit of the channel. This does lead to an unphysical artefact for small NWP volumes at the entrance, however: the NWP forms a hemisphere of small radius until the NWP volume is large enough that it touches the solid walls of the channel. This in turn implies a large capillary pressure before NWP reaches the throat. Therefore, in the results below, we show only calculations after the NWP volume is large enough to press against the solid walls at the entrance.

Surface Evolver then allows us to calculate the capillary pressure, the spatial distribution of fluid phases, and the stability of the interfaces. We fix the volume of WP and incrementally raise or lower it to simulate quasi-static motion of the NWP along a channel. Symmetry can be used to speed up the calculations: for example it is only necessary to simulate one quarter of the straight channel since all four corners are equivalent. 

To determine the stability of the interface, we examine the eigenvalues of the Hessian matrix of the energy~\cite{brakke96}: a change in the sign of the smallest eigenvalue to a negative value signals instability, and the software includes tools that allow us to determine the corresponding eigenmode. In particular, this is the case when two interfaces meet in a straight channel and, as we shall see below, when merging occurs in other channel geometries.

In a straight channel with a rectangular cross-section (figure~\ref{fig:straight}) there are side walls at $x = \pm \frac{1}{2} W$, with the bottom at $z=0$ and the top at $z=H$, the entrance at $y=0$ and exit at $y=L$. A channel length of $L=4$ is sufficient for a steadily propagating interface to form. The side walls are replaced by parts of a cylinder of radius $R_{curv}\pm \frac{1}{2}W$ to generate a curved rectangular duct.  Finally, cylindrical pillars with radius $R$ are placed at $x = \pm (\frac{1}{2}W+R)$, $y=0$ between a flat base to the channel at $z=0$ and a flat top at $z=H$.

Our simulations start by introducing NWP at $y=0$ into a channel that is initially full of WP. We reduce the WP volume incrementally until the capillary pressure $p_c$ exceeds the capillary entry pressure $p_c^e$. The capillary pressure then decreases as NWP invades the downstream body. If $p_c$ falls sufficiently during the invasion of the pore body, there can be ``Roof snap-off"~\cite{roof70} in the throat: WP flows back to the throat and blocks it~\cite{rossen03}.

For example, consider a meniscus of WP around the bases of two nearby cylindrical pillars (see figure~\ref{fig:simulated_network}(a)) with NWP everywhere else in the throat and filling the pore bodies on either side. At first the WP occupies only  the corners around the pillars, but as the WP volume $V$ increases these menisci spread outwards and the capillary pressure falls. We determine the point at which the two WP regions connect, as in figure~\ref{fig:isolated_pillars}(a) below, bridging the gap between the cylinders, and record the corresponding capillary pressure $p_c^{br}$. As its volume is increased further, WP rises up the sides of the pillars and the gap between the pillars is slowly filled with WP from below. When the WP reaches half-way up the pillars it meets the menisci coming down from above. At this point there is an instability (negative eigenvalue) associated with an eigenmode indicating contraction of the interface towards the centre of the channel. We identify this as snap-off~\cite{roof70,ransohoffgr87} and this value of the capillary pressure for snap-off as $p_c^{sn}$. We can take the Evolver calculations no further beyond this change in the topology of the interfaces.

\section{Results}

\subsection{Straight Channel}
\label{sec:straight}

\begin{figure}
\centerline{
(a)
\includegraphics[angle=0,width=0.45\textwidth]{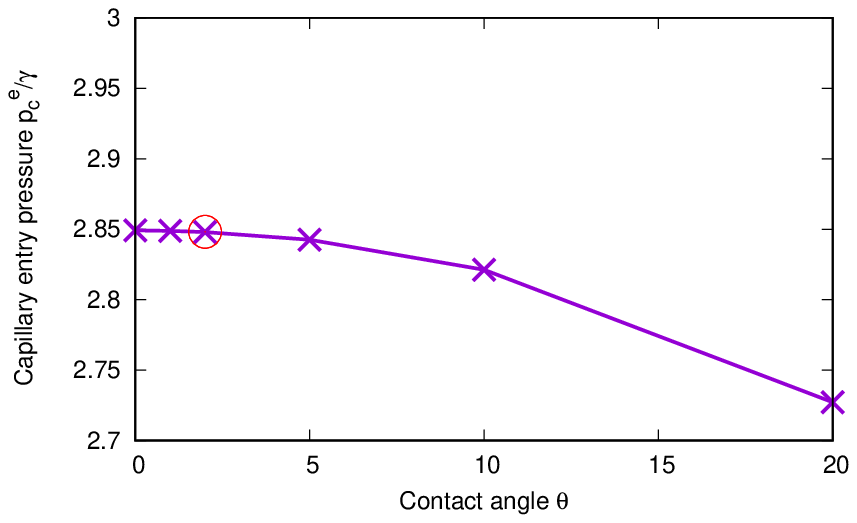}
(b)
\includegraphics[angle=0,width=0.45\textwidth]{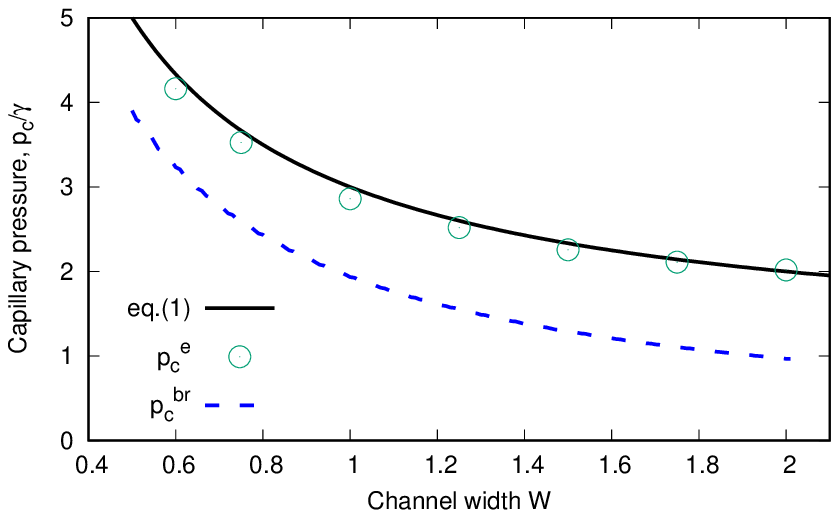}
}
\centerline{
(c)
\includegraphics[angle=0,width=0.45\textwidth]{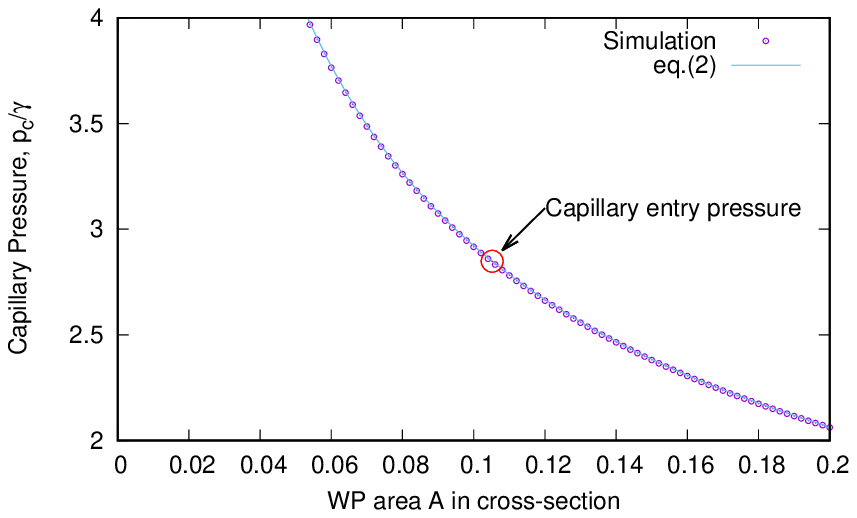}
}
\caption{Capillary pressure in straight rectangular channels. 
(a) Capillary entry pressure for different contact angles in a channel with $W=1$ and $H=2$. The value of $p_c$ for different contact angles is almost indistinguishable for $\theta$ less than about $5^\circ$, justifying our use of a small finite contact angle. The steady advance of the interface into the channel occurs with capillary pressure slightly less than the prediction of eq.~(\ref{eq:lenormand}), $p_c^e = 3$. 
(b) Capillary entry pressure as a function of width $W$ with fixed height $H=2$ and contact angle $2^\circ$. The capillary entry pressure is slightly over-estimated by eq.~(\ref{eq:lenormand}), while the capillary pressure at which bridging (and consequent snap-off) occurs is lower. The capillary entry pressure is greatest for a tall, narrow channel. 
(c) Capillary pressure as a function of the total area of WP in the cross-section of the channel in the four narrow regions of WP that remain after the interface has fully penetrated the channel, with $W=1$, $H=2$ and $\theta =2^\circ$. The particular case of the capillary entry pressure (from parts (a) and (b)) is highlighted to indicate how the shape of the interface spanning the channel determines the shape of the narrow regions of WP. }
\label{fig:data_straight}
\end{figure}

Figure~\ref{fig:data_straight}(a) shows the capillary entry pressure in a straight rectangular channel with width $W$ and height $H=2$ as a function of contact angle. As noted above, a contact angle of less than $5^\circ$ gives a good approximation to the result for perfect wetting. This is just below the prediction of eq.~(\ref{eq:lenormand}) for the capillary entry pressure. As the NWP enters the channel the capillary pressure is at its highest, but then drops by a few percent as a steady motion along the channel is established. 

Figure~\ref{fig:data_straight}(b) shows that increasing the width $W$ of the channel (while keeping its height $H$ fixed) reduces the capillary entry pressure, as predicted by eq.~(\ref{eq:lenormand}). The deviation in the value of $p_c^e$ from the prediction of eq.~(\ref{eq:lenormand}) (\cite{lenormand})  increases slightly as the channel gets narrower. Bridging occurs in the straight channel when the pair of upper or lower regions of WP (cf. figure~\ref{fig:straight}(b)) meet in the middle of the channel. At this point the capillary pressure has fallen from $p_c^e$; its value also decreases with increasing channel width, because the volume of WP required for bridging is greater. Moreover, in a straight channel, the capillary pressure for bridging, $p_c^{br}$, is equivalent to the value for snap-off, $p_c^{sn}$, since as soon as the NWP no longer touches all four walls of the channel a Rayleigh-Plateau instability occurs and the NWP breaks up into bubbles.

Eq.~(\ref{eq:mamorrow}) gives the value of $p_c$ for any WP volume in a rectangular channel before bridging (since the aspect ratio is irrelevant before the WP regions meet). Figure~\ref{fig:data_straight}(c) shows that not only do our simulations agree with the prediction of eq.~(\ref{eq:mamorrow}), but that during the motion of the NWP along the channel the area of the WP regions in the corners is entirely determined by the shape of the front at the capillary entry pressure.

\subsection{Curved rectangular duct}
\label{sec:curved}

\begin{figure}
\centerline{
(a)
\includegraphics[angle=0,width=0.4\textwidth]{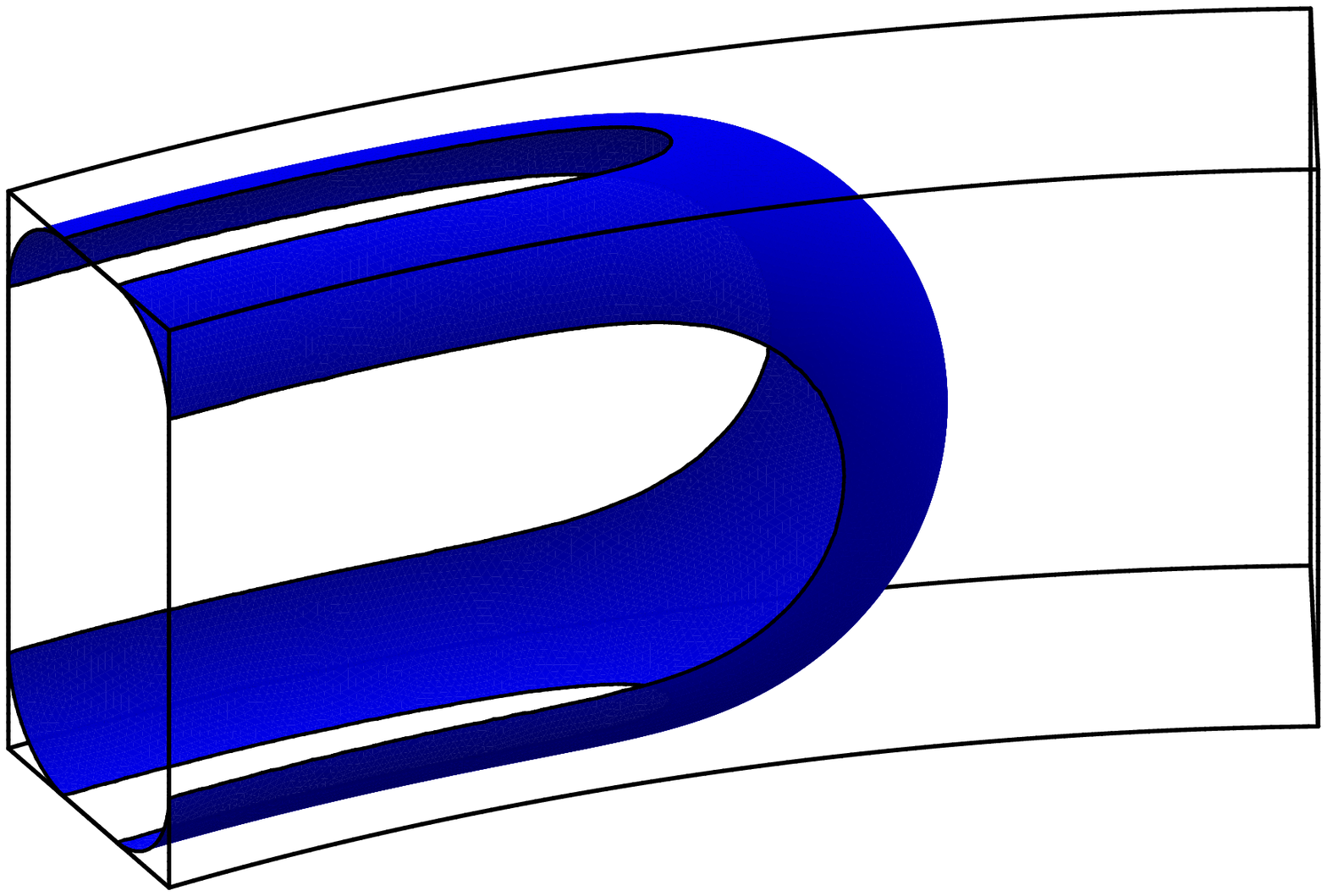}
(b)
\includegraphics[angle=0,width=0.55\textwidth]{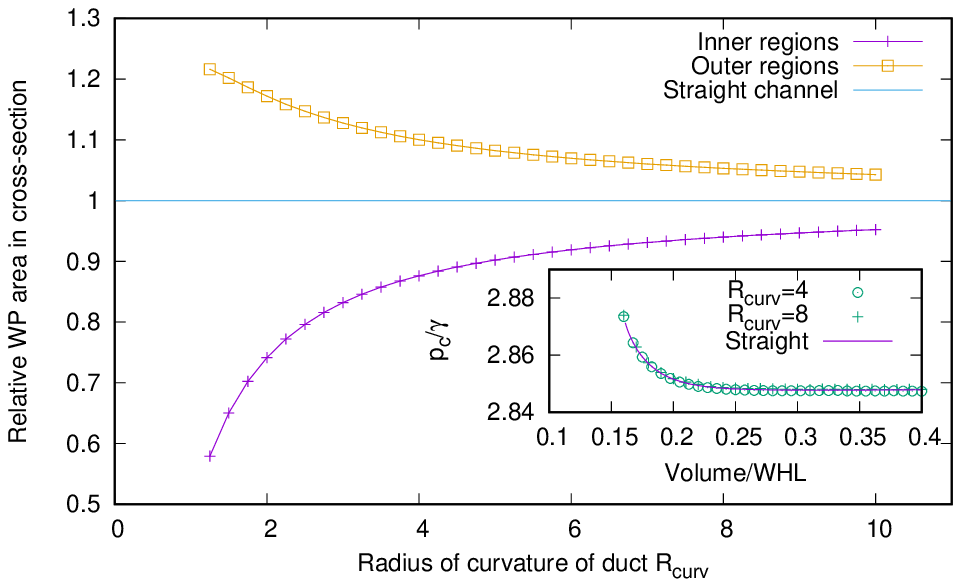}
}
\caption{(a) A curved rectangular duct with $H=2$, $W=1$ and $R_{curv}=8$, i.e. with curved walls having radius of curvature $R_{curv}\pm\frac{1}{2}W$. The interface separating the two fluids meets the walls at a contact angle of $2^\circ$. (b) Cross-sectional areas of the WP regions in the corners of a curved rectangular duct, normalized by the value for a straight channel. The inset shows that the capillary pressure is independent of the radius of curvature of the duct. The contact angle is $2^\circ$.}
\label{fig:data_curved}
\end{figure}

If the channel is a gently and uniformly curved rectangular duct with planar top and bottom, as shown in figure~\ref{fig:data_curved}(a), the capillary pressure of an interface moving along the duct is indistinguishable from its value in a straight channel. That is, the radius of curvature $R_{curv}$ of the centreline of the duct, which is bounded below by the half-width of the channel $W/2$, does not influence $p_c$.

\begin{figure}
\centerline{
\includegraphics[angle=0,width=0.65\textwidth]{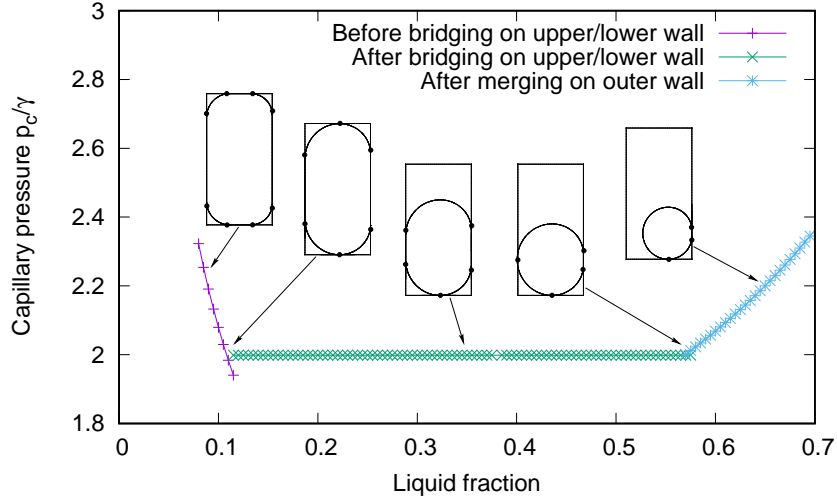}
}
\caption{Imbibition into a curved rectangular duct with radius of curvature $R_{curv} = 5$. The capillary pressure decreases until the first bridge is formed: if the channel height is greater than its width (here $H=2W$) this occurs first on the top (as shown) or bottom surfaces. The interface then remains in contact with three walls and the capillary pressure increases weakly until the merging transition on the outer wall.  Beyond this point the capillary pressure increases more rapidly. Also shown is the sequence of interface shapes in a vertical cross-section through the duct, with points at which the interface meets the duct surfaces highlighted with dots. The contact angle is $2^\circ$.  }
\label{fig:data_curved_pressures}
\end{figure}

However, the areas of the triangular regions that remain after the interface has passed are different on the inner and outer curved walls of the duct. Figure ~\ref{fig:data_curved}(b) shows that the cross-sectional area of these regions can deviate significantly from the straight channel case (in which the inner and outer regions are identical), with a greater deviation on the inner wall, but that they slowly converge towards the same value as the radius of curvature $R_{curv}$ increases. Their size will diverge more widely as the radius of curvature decreases if the volume of WP left in the duct after the front has passed is larger.

If, however, WP can re-imbibe into the channel then two or more of the four triangular regions may meet. There are two cases to consider. 
First, if the height $H$ is greater than the width $W$, as is the case in our ``standard" channel here, then bridging first occurs at the top and/or bottom of the duct, and they then merge on the outer wall, as shown in figure~\ref{fig:data_curved_pressures}.
Second, if the width $W$ is at least as great as the height $H$ (a short, wide, channel, including a duct with square cross-section), the order in which the transitions occur is reversed: the regions first merge on the outer wall, furthest from the centre of curvature, and then bridge across the top and/or bottom of the duct. We find (for contact angle $2^\circ$) that the the aspect ratio $H/W$ at which there is a transition from one case to the other, i.e. at which bridging and merging occur simultaneously on the upper/lower and outer surfaces, increases linearly, but weakly, with the curvature $1/R_{curv}$ of the duct, from $H/W=1$ for a straight channel to $H/W=1.2$ for $R_{curv}=2$.

In the first case, with $H=2$ and $W=1$, bridging across the top or bottom of the duct occurs at the same saturation (an area of WP in the cross-section of 0.12) and the same capillary pressure ($p_c \approx 1.9$)
for all values of the radius of curvature $R_{curv}$. That is, although the wetting phase is re-distributed between the triangular regions on the outer and inner sides of the duct as its curvature changes, and therefore the position at which they meet is different, the total volume of  WP and the capillary pressure at which the regions meet doesn't change.

Following bridging, there is an almost semicircular interface (in cross-section) moving vertically within the duct (figure~\ref{fig:data_curved_pressures}). Moreover, the capillary pressure increases very slightly with increasing volume of WP in this regime. Such a situation is unstable, as in the arguments of Roof~\cite{roof70} concerning the largest inscribed circle, and WP floods into the duct and blocks the flow of NWP. Thus the merging transition on the outer wall is never reached, and the capillary pressure for snap-off is the same as the capillary pressure for bridging: $p_c^{br} = p_c^{sn}$.

In the case of a square duct the situation is slightly different. The merging transition on the outer wall occurs at a capillary pressure that depends on the radius of curvature: it occurs at higher capillary pressures, and hence lower volumes of WP, in tightly-curved wide ducts. After merging the capillary pressure increases slightly, as for taller ducts, and so snap-off occurs and WP will flood into the duct.

\subsection{Gap between cylindrical pillars}
\label{sec:pillars}

Therefore, in both the straight and curved rectangular channels with height $H=2$ the capillary pressure at which bridging occurs is the same as the capillary pressure for snap-off. We now consider a situation in which the curvature of the side walls has a different sign on each side, and show that here this equivalence is {\em not} found.

\subsubsection{WP surrounding the base of two pillars}

\begin{figure}
\centerline{
(a)
\includegraphics[angle=0,width=0.5\textwidth]{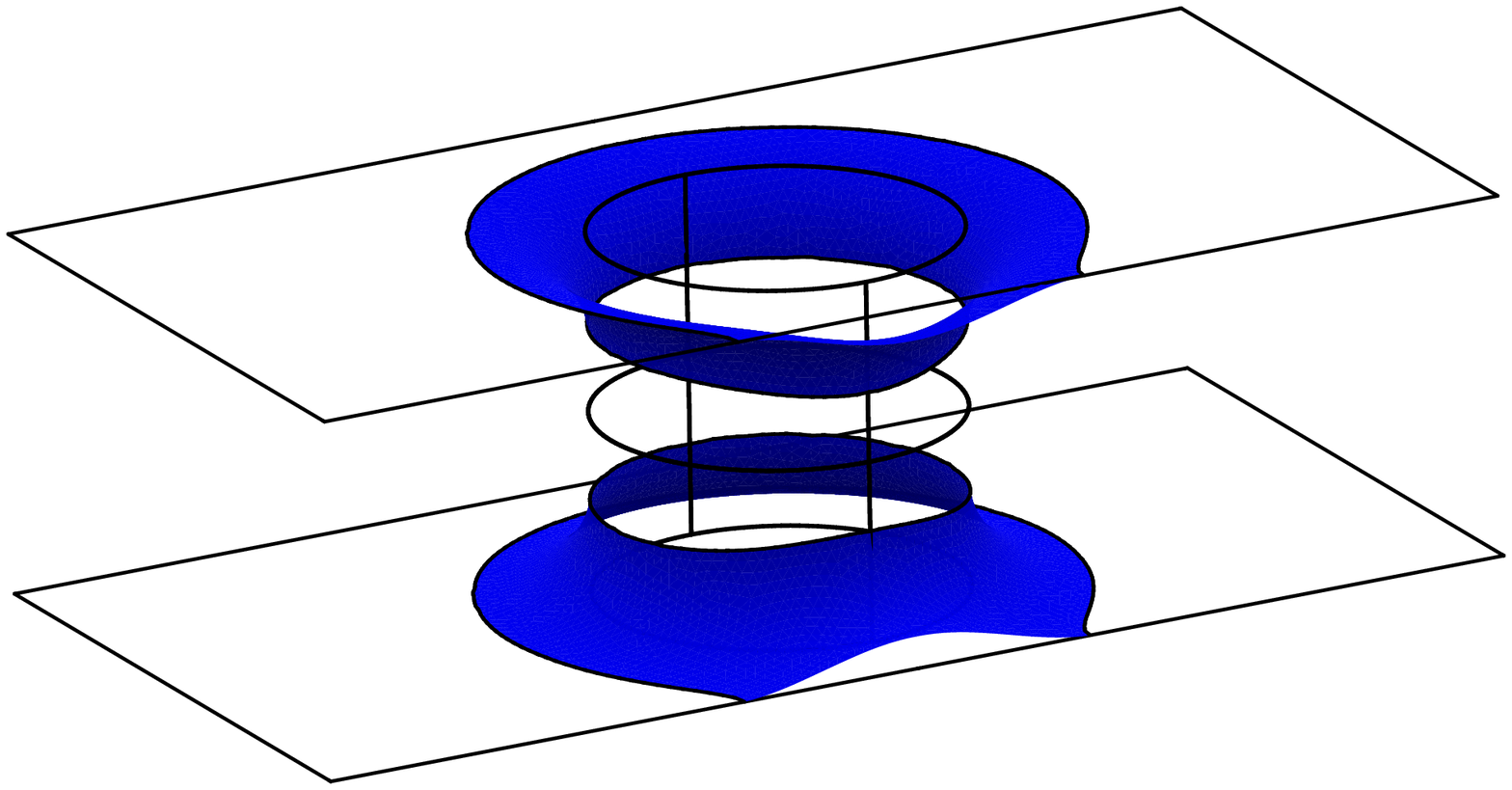}
(b)
\includegraphics[angle=0,width=0.5\textwidth]{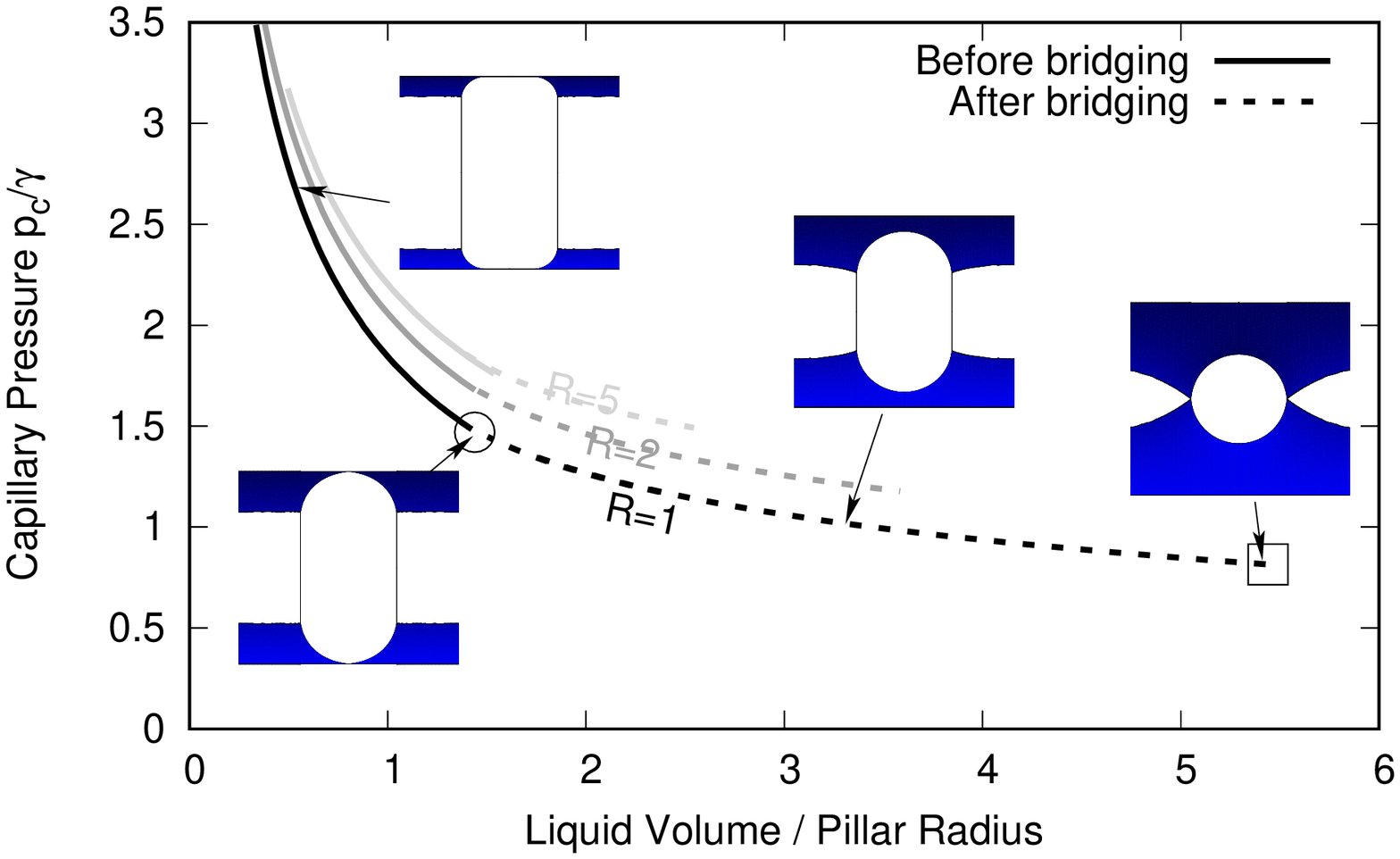}
}
\centerline{
(c)
\includegraphics[angle=0,width=0.5\textwidth]{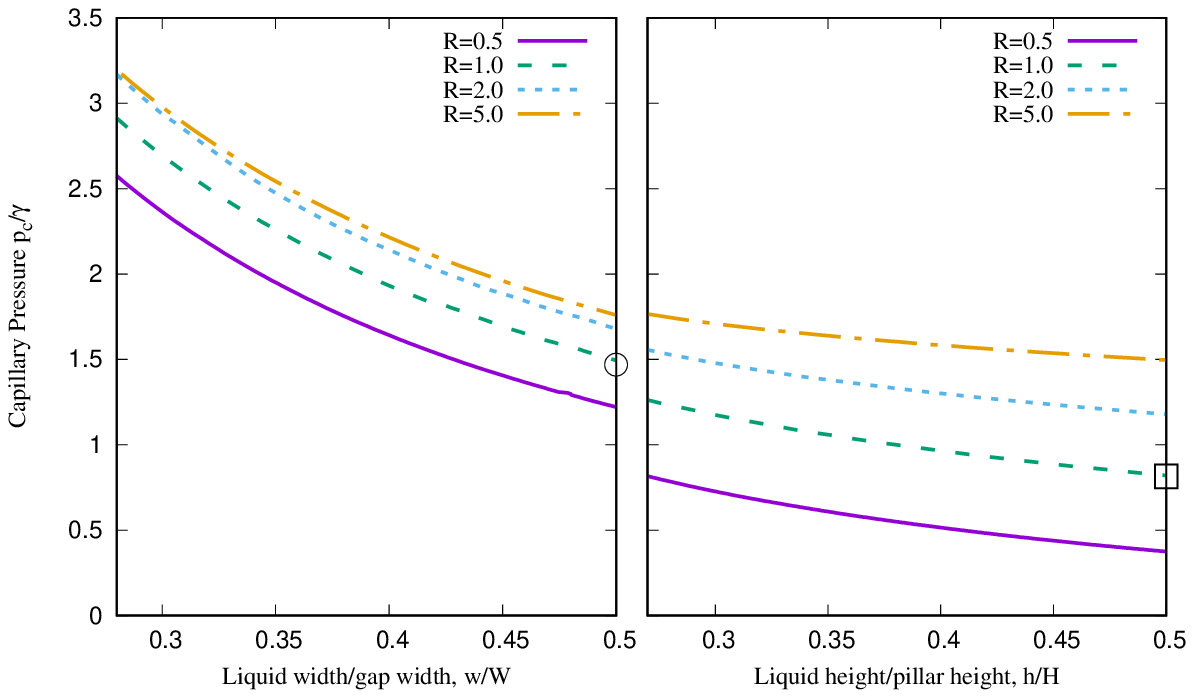}
(d)
\includegraphics[angle=0,width=0.5\textwidth]{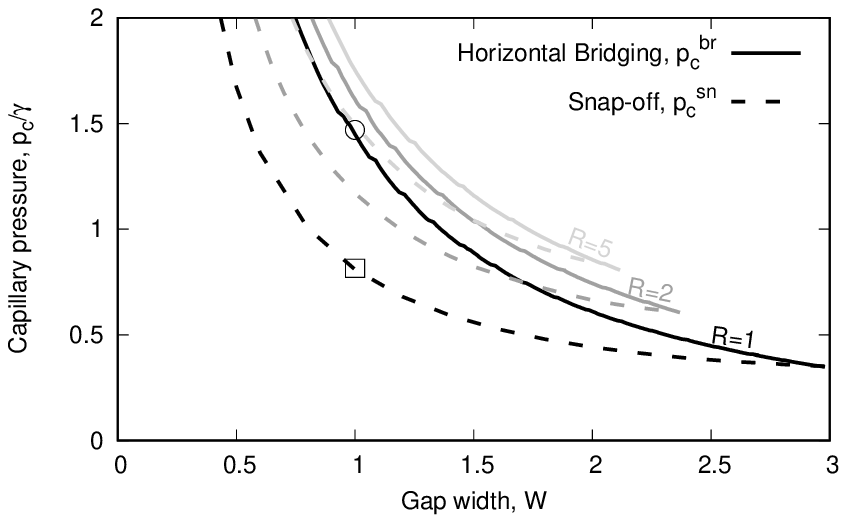}
}
\caption{Bridging and snap-off in a throat between cylindrical pillars. (a) Oblique view of WP around one of the two pillars, between the formation of the horizontal bridge and snap-off (subsequent merging), corresponding to $R=1$ and a total WP volume of 2.6. (b) Capillary pressure as a function of WP volume divided by radius, which is a measure of the cross-sectional area of the meniscus, for cylinders of radius $R= 1,2$ and $5$, with gap width $W=1$.  The circular symbol indicates bridging and the square symbol indicates snap-off here and in (c) and (d).  The images are views through the centre of the throat, with the pillars to either side, for $R=1$. (c)  The capillary pressure for different pillar radii $R$. When $w/W$ reaches 0.5 the WP bridges across the channel. When $h/H$ reaches 0.5 the two regions of WP touching the top and bottom surfaces merge and snap-off occurs.  (d) The critical capillary pressure at which bridging and snap-off occur as the gap width $W$ changes for pillars with radius $R=1, 2$ or $5$. Vertical merging occurs before bridging for values of gap width greater than the point at which the two curves meet for each $R$. Height is $H=2$ and the contact angle is $2^\circ$.}
\label{fig:isolated_pillars}
\end{figure}

As described above, we gradually increase the volume of WP around and in the gap between two cylinders of radius $R$ situated between parallel horizontal plates; one half (by symmetry) of the geometry is shown in figure~\ref{fig:isolated_pillars}(a). We fix the height to be $H=2$, as for the channels considered above, and vary the radius $R$  of the pillars and the gap $W$ between them to determine the different capillary pressures for bridging and snap-off.

It is instructive to first consider the shape of the interface surrounding an isolated pair of cylindrical pillars and its capillary pressure. Figure~\ref{fig:isolated_pillars} shows images of the interface and the capillary pressure.

The WP meniscus around the base and top of each pillar extends some distance $w$ across the gap and some height $h$ up the sides of the pillar. We use Surface Evolver to calculate the capillary pressure and the values of $w$ and $h$ as the WP volume $V$ varies for different gap widths $W$, as shown in figures~\ref{fig:isolated_pillars}(c) and (d). The capillary pressure decreases as the WP volume increases and $w$ and $h$ increase. When $w$ reaches $W/2$ the WP bridges the gap between the cylinders; we denote the critical capillary pressure at which this happens by $p_c^{br}$. Figure~\ref{fig:isolated_pillars}(b) shows that the volume of WP at which bridging occurs increases linearly with pillar radius $R$, but that $p_c^{br}$ increases only slightly with $R$.

Figure~\ref{fig:isolated_pillars}(c) indicates that the capillary pressure decreases further as more of the WP builds up in the gap between the pillars. There is a range of $p_c$ for which a bridge is stable without snap-off, corresponding to the difference in capillary pressure between the right hand sides of the two panels in figure~\ref{fig:isolated_pillars}(c) for each value of $R$. The upper and lower WP regions in the gap between the pillars meet, half-way up the pillars, when their height is $h=\frac{1}{2}H$, at which point we denote the critical capillary pressure by $p_c^{sn}$, the right-most values in figure~\ref{fig:isolated_pillars}(c). Note that at the point where the bridging between upper and lower WP regions occurs the cross-section of the interface appears circular; the gradient of $p_c(V)$ is very shallow here (figure~\ref{fig:isolated_pillars}(c), right hand panel), indicating that the WP has almost reached the critical point for snap-off, beyond which it floods back into the gap between the pillars~\cite{ransohoffgr87}. Our calculations and assessment of the eigenmodes given by Surface Evolver suggest that this instability occurs immediately as these two regions meet, and the interface changes topology.

At small gap widths there is a significant difference between the capillary pressure for bridging and the capillary pressure for snap-off, since bridging can happen at small WP volume, while the WP volume must build up significantly before vertical merging occurs. As the gap between the pillars widens (increasing $W$, figure~\ref{fig:isolated_pillars}(d)), more WP must accumulate before bridging occurs. Consequently, the capillary pressure for bridging decreases with increasing $W$. 

At large enough gap widths, above about $W=3$ in the case $R=1$, merging occurs before horizontal bridging; this gives rise to a capillary bridge connecting the top and bottom of the cylinder, but does not permit two-phase flow.

\subsubsection{Asymmetric case}

Up until now we have considered a situation in which the WP around both pillars is at the same capillary pressure. But it is also possible for WP to accumulate only in part of the pore network connected to the point of injection and it must then form bridges across pore throats to give it a connection to flow across the network.  

When NWP first crosses the pore network, as illustrated in figure~\ref{fig:simulated_network}(b), the connections between regions of WP are broken. As WP accumulates, the capillary pressure falls, but only in the interconnected network of WP-filled pores connecting back to the point of injection.  Therefore, when these critical bridges form (i.e., those that give WP connection across the network), one pillar is connected to the reservoir of WP at low capillary pressure while the other pillar is not. The bridge has to form with liquid initially accumulating on one side only.

What happens after the bridge forms?  Clearly, the dynamics of bridge formation are complex in the short term, and we do not consider them here. But the result is that both sides are now at the lower value of capillary pressure associated with the WP's flowing network.  So the question is: what is $p_c$ on the wet pillar at the point when the bridge forms?  Is it low enough that, once this value of $p_c$ is established on both sides, there would be snap-off without a bridge?

To illustrate the idea, we continue to take a gap width of $W=1$, pillars of radius $R=1$, and assume that the WP meniscus surrounding the dry pillar has capillary pressure $p_c \approx 3$. From the left hand side of figure~\ref{fig:isolated_pillars}(c) we see that the WP extends a distance slightly less than $0.3W$ from the base of the pillar in this case. Bridging will therefore occur if the WP around the wet pillar extends to a distance of $0.7W$ from that pillar. The capillary pressure around the wet pillar at bridging is the value of $p_c^{br}$ for a gap of $W=1.4$, which we read from figure~\ref{fig:isolated_pillars}(d) as $p_c/\gamma \approx 1.0$. This is {\em greater} than the capillary pressure for snap-off (figure~\ref{fig:isolated_pillars}(b)), $p_c/\gamma \approx 0.8$, and hence bridging will not lead to snap-off in this case.

For snap-off to occur immediately following bridging in this asymmetric case therefore requires that the extent of the WP meniscus around the dry pillar is smaller than hypothesised, and the WP at higher capillary pressure.
Nonetheless, capillary pressure could be greater on the dry pillar than the capillary entry pressure for the throat, as follows. As gas invades the pore network, as shown in figure~\ref{fig:simulated_network}(b), capillary pressure rises throughout the network according to the capillary entry pressure of the latest throat to be invaded. As noted above, that process stops when gas forms a continuous path across the network and liquid continuity across the network is broken. Thus the capillary pressure on the dry  pillar is at the capillary entry pressure of the last throat to be invaded by gas, not that of the given throat.  Figure~\ref{fig:isolated_pillars}(d) indicates that the capillary pressure for snap-off with $W=1$ (again with $R=1$) is equal to the capillary pressure for bridging at a gap width of $W \approx 1.7$. So only when the WP meniscus around the dry pillar extends no more than 15\% of the way across the gap, with a capillary pressure of $p_c/\gamma \approx 5$, far in excess of the likely capillary entry pressure, is snap-off without bridging possible.

However, for larger pillar radii the curves of $p_c$ against gap width are less steep (figure~\ref{fig:isolated_pillars}(d)), the difference between $p_c^{br}$ and $p_c^{sn}$ smaller (figure~\ref{fig:isolated_pillars}(c)), and so this transition to snap-off in the asymmetric case becomes more likely and flow without fluctuating pore occupancy less likely. For example, with $R=2$ the value of $p_c$ drops from its value for bridging at $W=1$ to below the value for snap-off at $W=1$ when $W \approx 1.4$, suggesting that snap-off would immediately follow bridging driven from one side. 

For larger gap-widths it is more likely that vertical merging would occur before bridging between the pillars, something which our calculations do not currently resolve.

\subsubsection{Invasion of NWP into a WP-filled gap between pillars}

\begin{figure}
\centerline{
(a)
\includegraphics[angle=0,width=0.45\textwidth]{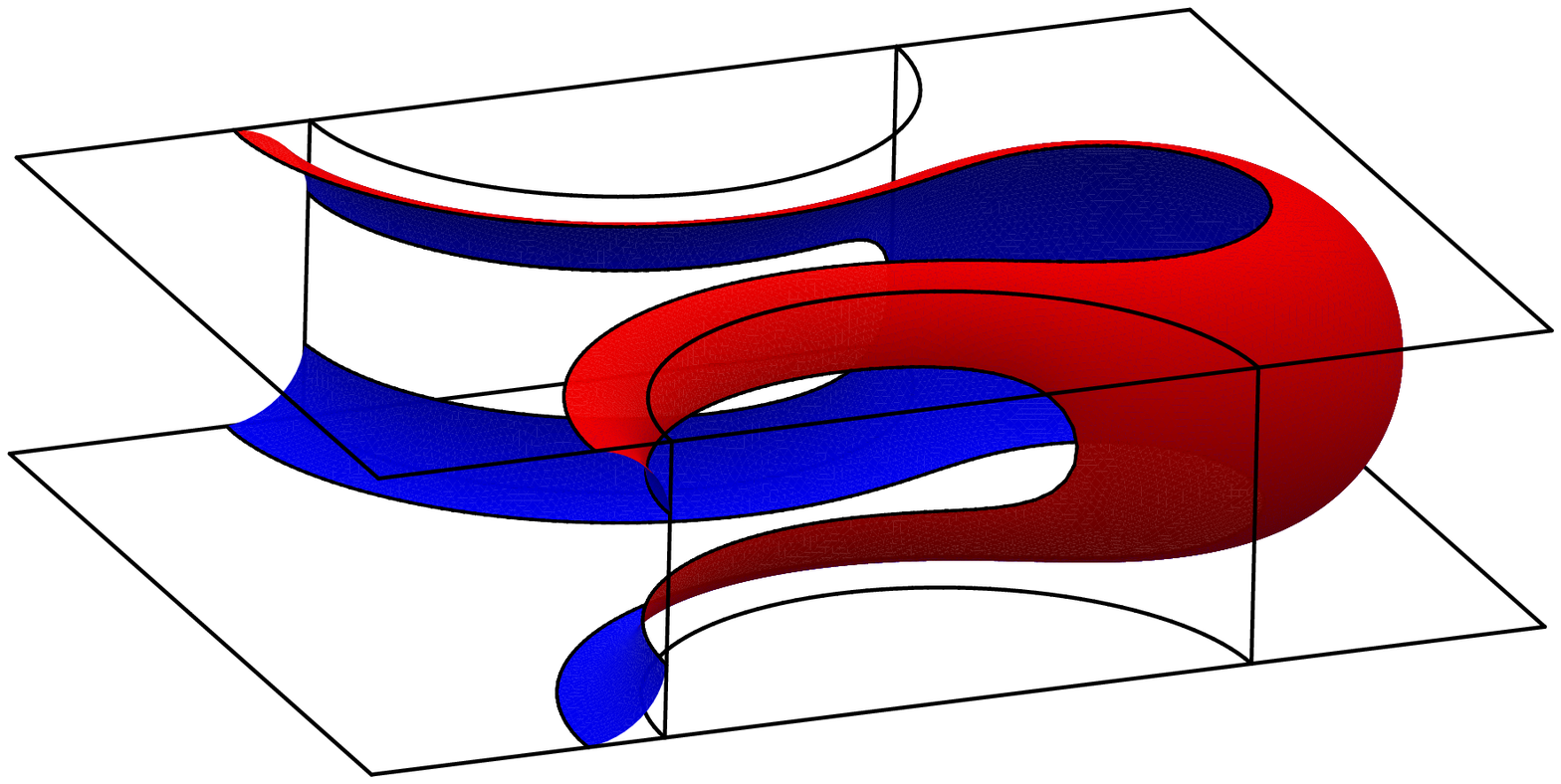}
(b)
\includegraphics[angle=0,width=0.45\textwidth]{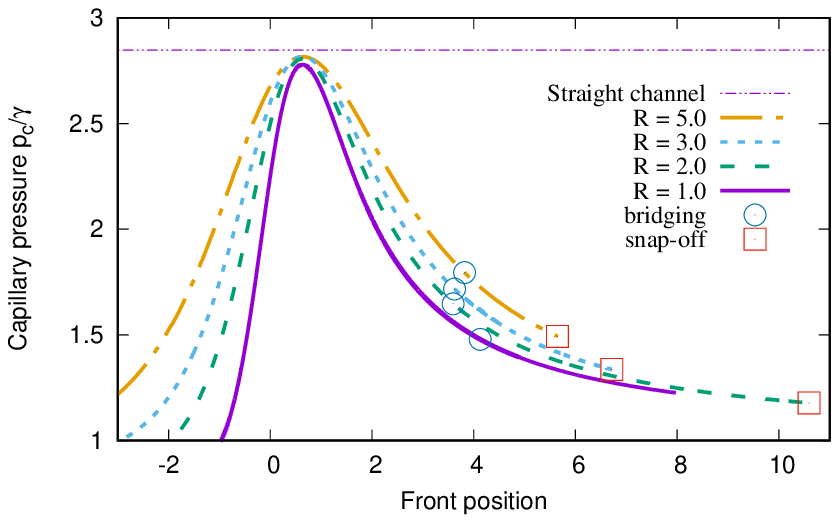}
}
\centerline{
(c)
\includegraphics[angle=0,width=0.45\textwidth]{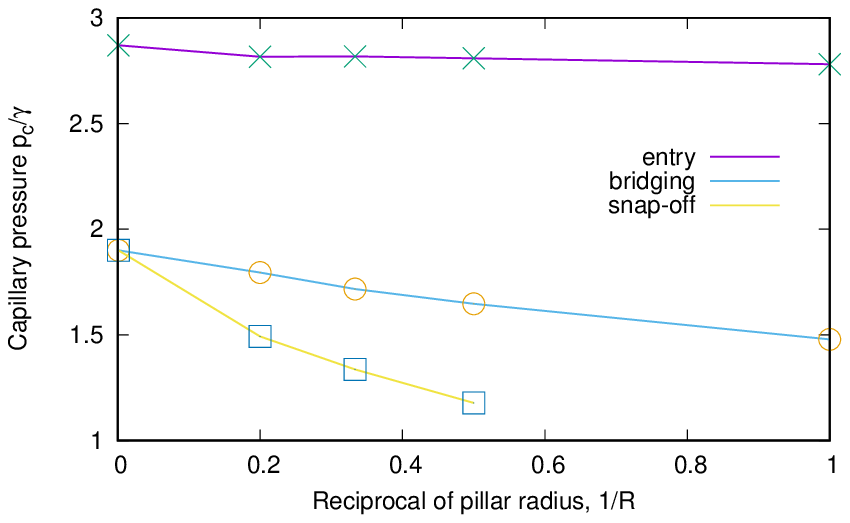}
}
\caption{NWP invasion of the pore throat between pillars.
(a) The interface moves from left to right through the gap between two cylindrical pillars, leaving the displaced WP  around the base and top of each pillar. Pillar radius is $R=1$ in this case.  
(b) The maximum capillary pressure occurs just after the leading edge of the front has passed the centre of the gap. It increases slightly with the radius of the pillars. Also shown is the predicted capillary pressure in a straight channel, slightly below the prediction of $p_c^e=3$ from eq.~(\ref{eq:lenormand}).   
(c) A summary of the different critical capillary pressures, plotted as a function of $1/R$ so that the values for the straight channel are shown on the left. 
Gap width is $W = 1$, height is $H=2$, and the contact angle is $2^\circ$.}
\label{fig:pillars}
\end{figure}

Snap-off triggered by NWP invasion of the downstream pore body is called ``Roof" snap-off~\cite{roof70, rossen03,ransohoffgr87}. We turn now to the invasion of NWP into a WP-filled gap between the pillars, i.e. during initial drainage of the pore network. We fix the gap width to be $W=1$, much less than the height $H$.  NWP moves into the gap much as in the case of a rectangular channel (\S~\ref{sec:straight}), with a roughly hemi-spherical front (figure~\ref{fig:pillars}(a)) and leaving behind narrow regions of WP in the corners of the channel where the pillars meet the upper and lower surfaces, which correspond to the images in figure~\ref{fig:isolated_pillars}(b) at high capillary pressure.

Figure~\ref{fig:pillars}(b) shows the capillary pressure across the interface as it moves through the throat for four different values of the pillar radius $R$. The capillary pressure is greatest for the widest pillars, and for sufficiently large $R$ approaches the value for a straight channel. The maximum capillary pressure (i.e. the capillary entry pressure $p_c^e$) occurs just {\em after} the front (measured as the leading position on the interface along the axis through the gap) has passed between the pillars, and this asymmetry is also apparent in the steepness of the pressure curves before and after this point.

After the front has passed through the gap between the pillars, the capillary pressure decreases as the NWP spreads out into the pore space beyond. The narrow regions of WP in the corners of the channel shrink and swell as the capillary pressure increases and then decreases as the interface invades the pore throat and then the downstream pore body~\cite{video}.

We can estimate the horizontal extent of the interface downstream, beyond the pillars as follows: far downstream of the pillars the interface has a semi-circular vertical cross-section (in the perfect wetting case with contact angle $\theta = 0$), spanning the gap between the flat upper and lower surfaces (see figure~\ref{fig:pillars}(a)). If we also approximate its horizontal cross-section as a circle of diameter $D$, then the capillary pressure here is $p_c = \gamma \left(\frac{2}{H} + \frac{2}{D}\right)$, similar in spirit to the derivation of eq.(~\ref{eq:lenormand}). 

\begin{figure}
\centerline{
(a)
\includegraphics[angle=0,width=0.45\textwidth]{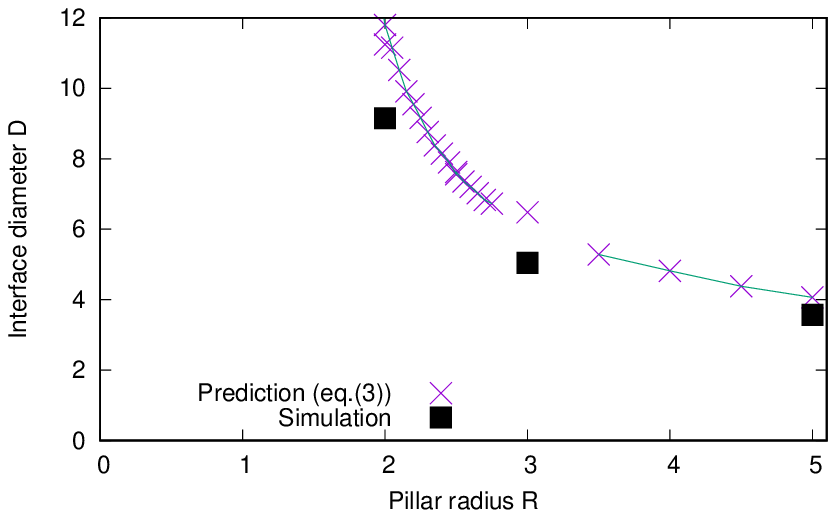}
(b)
\includegraphics[angle=0,width=0.45\textwidth]{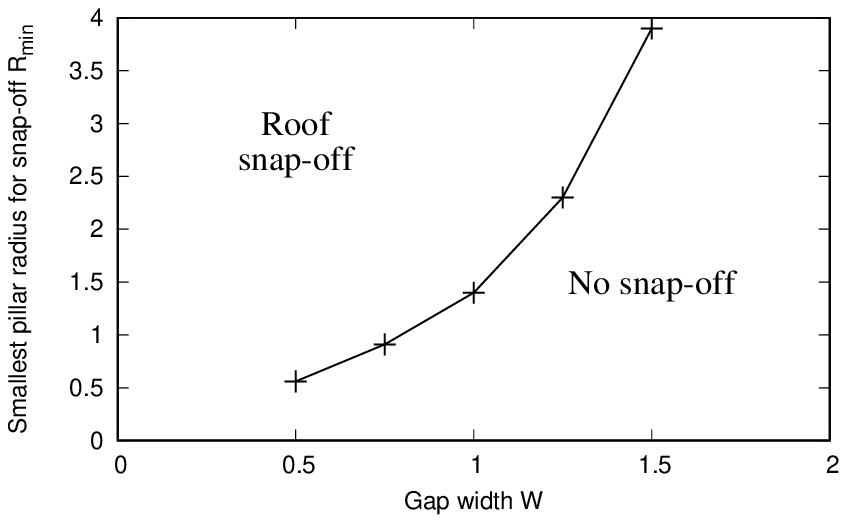}
}
\caption{
(a) The diameter of the bubble that is created in the pore body downstream when snap-off occurs in the gap between the pillars after the interface has passed: comparison of eq.~(\ref{eq:Dpred}) using capillary pressure values from figure~\ref{fig:isolated_pillars} with direct simulation as in figure~\ref{fig:pillars}(a). Gap width is $W = 1$. 
(b) The minimum pillar radius for snap-off in a gap of given width. The region below the curve is inaccessible because the capillary pressure for snap-off is below $2 \gamma/H$.
Height is $H=2$, and the contact angle is $2^\circ$.}
\label{fig:bubblesize}
\end{figure}

The interface must have the same capillary pressure everywhere, so we compare this expression for the capillary pressure to the critical value $p_c^{sn}$ for snap-off in the gap between the pillars when $h=H/2$. We extrapolate the data shown in figure~\ref{fig:isolated_pillars}(c) for $p_c^{sn}(h=\frac{1}{2}H)$ to more values of $R$ and then estimate the minimum pore-body diameter at which snap-off can occur in the pore throat:  
\begin{equation}
D = 2/(p_c^{sn}/\gamma - 2/H).
\label{eq:Dpred}
\end{equation}
This is shown in figure~\ref{fig:bubblesize}(a), where it is compared with direct measurements of the width of the interface from simulations, such as the one shown in figure~\ref{fig:pillars}(a), that have been allowed to run until the capillary pressure decreases down to $p_c^{sn}$ and the WP fills the gap between the pillars as a result of Roof snap-off. 

For gently constricted pores with circular cross-section along the flow axis the criterion for Roof snap-off is that the diameter of the pore body be at least twice the diameter of the throat. In this case, with $W=1$, that suggests $D=2$~\cite{roof70}. In microfluidic devices of uniform depth, pore bodies must be wider than this to trigger Roof snap-off. Strongly-convex throats, reflected here in small $R$, stabilize bridging against snap-off during subsequent WP imbibition, but may not allow Roof snap-off during initial drainage of the network. The value of $D$ that we predict here is close to the direct measurement shown by the black squares in figure~\ref{fig:bubblesize}(a); it is a small multiple of the gap width for pillars with large radius, but rises rapidly for smaller pillars, and tends to infinity somewhere between $R=1$ and $R=2$. With $H=2$ we see that $D \rightarrow \infty$ for $p_c^{sn}/\gamma \rightarrow 1$, which figure~\ref{fig:isolated_pillars}(d) indicates is indeed the case for $R$ between 1 and 2. In all cases it is much larger than the criterion for pore bodies and throats of circular cross-section.

Hence for sufficiently tightly-curved pillars (highly-convex pore throats), Roof snap-off does {\em not} occur during NWP invasion of the pore throat however wide the pore body. The minimum possible pillar radius $R_{min}$ at which $D \rightarrow \infty$ also depends on the gap width $W$. By calculating the cylinder radius at which, for a given gap width $W$, the capillary pressure for snap-off is equal to $2\gamma/H$ (which is equal to one here), we can determine $R_{min}(W)$, as shown in figure~\ref{fig:bubblesize}(b). If the gap is narrow then Roof snap-off can occur in a sufficiently wide pore body;  as the gap widens the minimum pillar radius grows rapidly.

\section{Discussion: Implications for Two-Phase Flow in Microfluidics}

Our findings can be summarized as follows: bridging is not possible without snap-off in straight or curved ducts. Bridging is possible in tightly concave throats, represented here by throats between narrow-radius pillars. In that geometry, as shown in figure~\ref{fig:isolated_pillars}, there is a range of capillary pressure at which bridging is stable without snap-off. For very tightly-concave throats, however, Roof snap-off may not be possible, no matter how wide the downstream pore body (figure~\ref{fig:bubblesize}).  In all cases the width of a pore body in comparison to the width of a throat, as viewed from above, that would give Roof snap-off is much greater than that for pores and throats of circular cross-section~\cite{roof70}.

We have examined the feasibility of bridging in various pore-throat geometries. Of the examples shown in figure~\ref{fig:network}, (a) and (b) feature straight, rectangular ducts; most throats in (c) approximate pillars of very large radius; those in (d) include curved rectangular ducts, straight channels and some tightly-concave throats. Based on these geometries, we conclude that it is unlikely that two-phase flow is possible in the networks shown in figure~\ref{fig:network}, or similar networks, without fluctuating pore occupancy (although it might be possible to design a network that would have the right range of throat geometries for such fluctuations to be suppressed). In three-dimensional geological pore networks, two-phase flow without fluctuating pore occupancy can occur at arbitrarily small pressure gradients~\cite{kingm19,rossen03}. At higher pressure gradients, quantified through the dimensionless capillary number, flow with fluctuating pore occupancies and displacement of trapped, isolated NWP is possible~\cite{lakejrp14,tangsvr19}. The flow regime above this transition in capillary number differs significantly from that below it~\cite{lakejrp14}. This suggests that imposed two-phase flow in most current microfluidic devices must necessarily reflect the high-capillary-number flow regime.

Steady two-phase flow is possible in networks featuring concave pore throats, especially tightly-concave pore throats. As illustrated in figure~\ref{fig:simulated_network}(b), under conditions where both phases might flow, the flow path for each phase is extremely inefficient. That suggests that the relative permeability of each phase~\cite{kingm19,lakejrp14} is very low. One implication is that even in conditions where steady two-phase flow is possible, the pressure gradient required to sustain the imposed flow rate could push the capillary number above the point where pore occupancy fluctuates. This is the subject of ongoing research that will be reported separately.

\end{document}